\documentclass[]{elsarticle}

\usepackage{amssymb}

\usepackage{float}
\restylefloat{table}

\usepackage{subcaption}

\RequirePackage{doi}
\usepackage{hyperref}
\hypersetup{
colorlinks=true,
}
\usepackage{natbib}

\clearpage{}\newcommand{\advligorts}{\texttt{advligorts}}
\newcommand{\guardian}{\texttt{guardian}}
\newcommand{\EPICS}{\texttt{EPICS}}
\newcommand{\MATLAB}{\textsc{Matlab}}

\clearpage{}

\journal{SoftwareX}

\begin{document}

\begin{frontmatter}

\title{\advligorts: The Advanced LIGO Real-Time Digital Control and
  Data Acquisition System}

\author[cit]{Rolf Bork}
\author[lho,cit]{Jonathan Hanks}
\author[lho,cit]{David Barker}
\author[llo,cit]{Joseph Betzwieser}
\author[cit]{Jameson Rollins}
\author[llo,cit]{Keith Thorne}
\author[lho,cit]{\mbox{Erik von Reis}}

\address[cit]{California Institute of Technology, Pasadena, CA 91125, USA}
\address[lho]{LIGO Hanford Observatory, Richland, WA 99352, USA}
\address[llo]{LIGO Livingston Observatory, Livingston, LA 70754, USA}

\begin{abstract}
  The Advanced LIGO detectors are sophisticated opto-mechanical
  devices.  At the core of their operation is feedback control.  The
  Advanced LIGO project developed a custom digital control and data
  acquisition system to handle the unique needs of this new breed of
  astronomical detector.  The \advligorts\ is the software component
  of this system.  This highly modular and extensible system has
  enabled the unprecedented performance of the LIGO instruments, and
  has been a vital component in the direct detection of gravitational
  waves.
\end{abstract}

\begin{keyword}

  real-time processing \sep feedback control \sep hardware control \sep data acquisition

\end{keyword}

\end{frontmatter}

\begin{table}[H]
\begin{tabular}{|l|p{5.0cm}|p{8.0cm}|}
\hline
\textbf{Nr.} & \textbf{Code metadata} & {} \\
\hline
C1 & Current code version & \textbf{4.0\~{}pre} \\
\hline
C2 & Permanent link to code/repository used for this code version & \url{https://git.ligo.org/cds/advligorts} \\
\hline
C3 & Code Ocean compute capsule & n/a \\
\hline
C4 & Legal Code License & GPLv3 \\
\hline
C5 & Code versioning system & git \\
\hline
C6 & Software code languages, required tools and services & C/C++, Perl, Python \\
\hline
C7 & Compilation requirements, operating environments & C99, Linux OS \\
\hline
C8 & Developer documentation/manual & \url{https://dcc.ligo.org/LIGO-E1200653} \\
\hline
C9 & Support email & \href{mailto:jameson.rollins@ligo.org}{jameson.rollins@ligo.org} \\
\hline
\end{tabular}
\caption{Code metadata}
\end{table}

\newpage

\section{Motivation and significance}

The development of long baseline gravitational wave detectors over the
last 40 years has been of groundbreaking scientific impact.  These
sophisticated measurement devices are revolutionizing astronomy and
astrophysics by providing an entirely new perception of the universe.

The complexity and scale of the LIGO detectors was novel.  While
prototype interferometer gravitational wave detectors paved the way,
their relative simplicity could not provide a clear roadmap for the
ultimate design of the LIGO detectors.  This was particularly true for
the interferometer feedback control systems at the core of the
instruments' operation.

Feedback control enables the LIGO interferometers to maintain the
operating point of a highly nonlinear precision optical instrument.
Sensors provide information about the state of the interferometers
that are fed as error signals to feedback controllers that filter and
transform the signals to produce control signals that are fed back to
the instrument to actuate on its state.  The dynamics of the
instruments were extensively modeled before their construction, but
the final design of the control loops could not be fully conceived a
priori.  The feedback controllers needed to be flexible to account for
uncertainty in the final feedback control scheme.  This motivated a
critical decision early in the LIGO design process: to use digital
instead of analog feedback control.

At the time, it was unclear if digital control was feasible for this
application.  There were concerns that digital control loops would
have insufficient bandwidth to control the instrument, or that the
analog$\leftrightarrow$digital conversion processes would inject too
much noise into the feedback loops, limiting the sensitivity of the
detectors.  Digital communication delays over the four kilometer arm
length was also a concern.  However, analog electronics are difficult
and time consuming to modify, which
would have severely limited the rate at which changes to the
controllers could have been made and, therefore, the rate of
refinement of the instruments.  Systems that could have allowed for
faster turnaround -- by making any potentially desirable parameter
changes easily switchable -- would have necessarily been complex,
difficult to design and maintain, and expensive.

The parameters of digital controllers, on the other hand, can be
modified nearly instantaneously, allowing much faster refinement.
Further modeling also made clear that excess noise from the
digitization and analogization processes could be sufficiently
mitigated by proper design of analog signal conditioning, and by
expected improvements to the bit depth and noise performance of
analog$\leftrightarrow$digital conversion electronics.  The increasing
speed of computers would allow the bandwidth and complexity of the
controllers to steadily increase over time.  Ultimately, it was
determined that the ease of quick modification - aided by the
promising advancements in performance - made digital feedback the
necessary choice.

The digital controllers for the Initial LIGO
detectors~\cite{doi:10.1088/0034-4885/72/7/076901} were handwritten C
code that was compiled into dedicated real-time Linux kernels.
Modular filter banks that allowed for in situ switching of feedback
filters and gains were inserted at key points in the feedback path to
allow instrument scientists to change the controller response on the
fly.  The feedback logic itself was hard-coded and difficult to
modify.  This was sufficient for Initial LIGO, with its small number
of feedback control loops.  Advanced LIGO was significantly more
complicated and required digital feedback controllers that were more
flexible.

For Advanced LIGO~\cite{doi:10.1088/0264-9381/32/7/074001}, a more
modular, extensible, and usable system was designed from the ground
up.  The \advligorts\ system gave scientists the ability to visually
represent the feedback signal flow and control logic in a more
intuitive way using a graphical user interface.  The signal flow
diagrams could be compiled into real-time code and re-loaded into the
front end computers in a matter of minutes, considerably speeding up
the turn around time to affect changes on the system.

The \advligorts\ system helped bring the new gravitational wave
detectors to fruition, enabling the first detection of gravitational
waves in 2015~\cite{doi:10.1103/physrevlett.116.061102}.

\begin{figure}
  \includegraphics[width=\columnwidth]{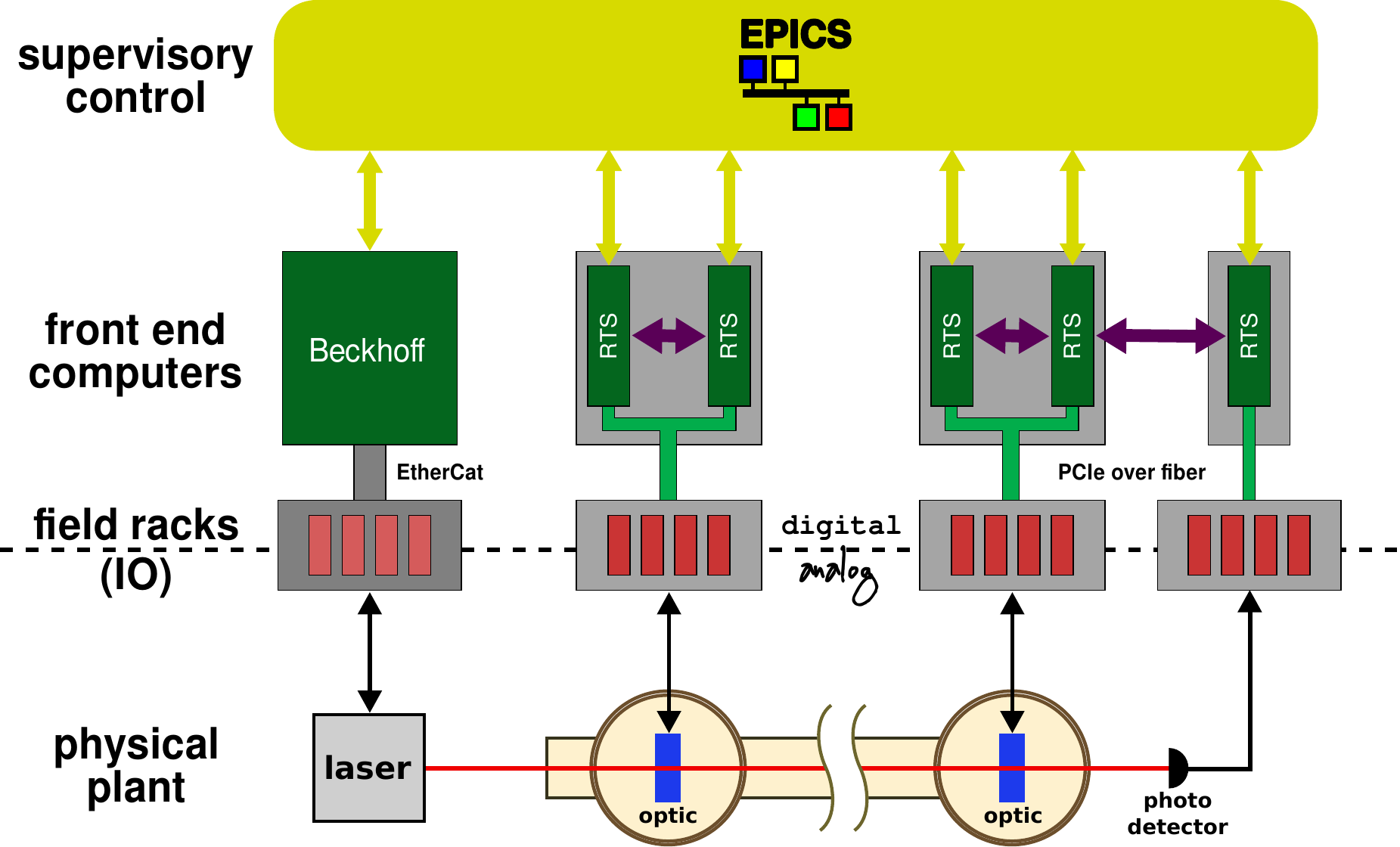}
  \caption{Overview of Advanced LIGO's digital instrument control
    system.  The physical plant consists of the interferometers
    themselves, and their sensors and actuators.  Signals are
    digitized and interpolated in the field racks, which are
    physically separated from the front end computers and connected to
    them via PCIe fiber.  The front end computers handle all real-time
    control with the \advligorts\ software (RTS).  Supervisory control
    and operator interfaces communicate with the RTS via the
    \EPICS\ message passing interface.  Beckhoff is a commercial
    PC-based programmable logic controller used for some slow control
    tasks.  Beckhoff uses the EtherCAT protocol to communicate with
    hardware devices and \EPICS\ to communicate with the rest of the
    system.}
  \label{fig:over}
\end{figure}

\section{Software description}

\begin{figure}
  \includegraphics[width=\columnwidth]{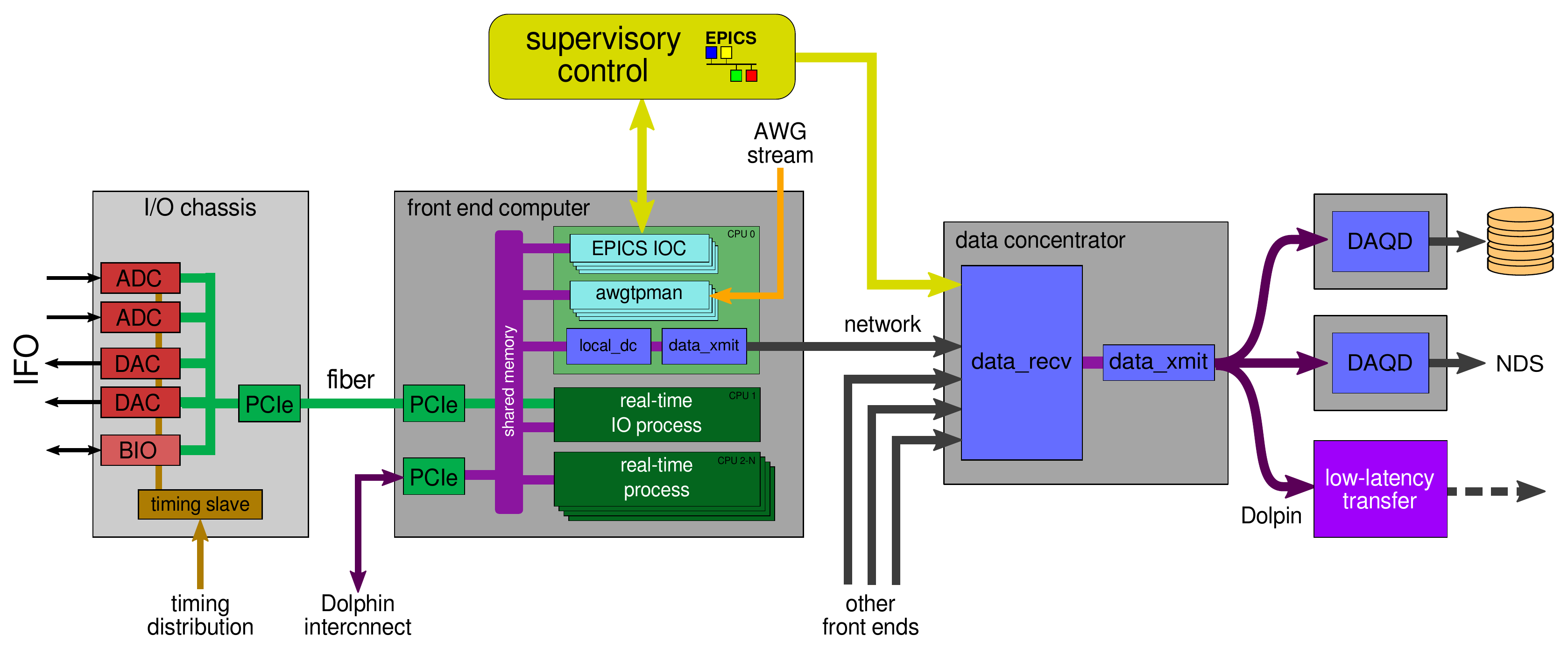}
  \caption{Architecture of the Advanced LIGO RTS.  At left is the I/O
    chassis that holds ADC, DAC and binary I/O (BIO) cards, and the
    timing slave that clocks the hardware.  The I/O chassis is
    connected to the front end host computer via a fiber PCIe bus
    extension.  The front end computer runs an RTS version of the
    Linux kernel, into which multiple RCG-generated RTS kernel modules
    (dark green) can be loaded.  The user space RTS processes are
    shown in cyan.  The data acquisition components are shown in blue.
    The ``data concentrator'' collects data from the distributed front
    end computers and passes the concatenated data to a DAQD process
    that writes it to disk, a DAQD network data server (NDS), or to
    the low-latency streaming service.}
  \label{fig:rts}
\end{figure}

\advligorts\ is the software component of the full Advanced LIGO
digital control and data acquisition system (hereafter also referred
to as the ``real-time system'' or ``RTS'').  The hardware consists of
analog-to-digital (ADC) and digital-to-analog (DAC) converters, binary
I/O modules, a timing distribution system to clock the ADCs and DACs,
and PCIe buses that interface all the hardware to the front end host
computers that run the \advligorts\ software (see
Figure~\ref{fig:over}).  The \advligorts\ software reads from and
writes to the hardware, executes all the digital control logic, and
passes data to the data acquisition pipeline.  Figure~\ref{fig:rts}
shows a more detailed schematic of this architecture, showing both
hardware and software components.

A important feature of \advligorts\ is its use of the Experimental
Physics and Industrial Control System (\EPICS), a Free and open-source
message passing system.  \EPICS\ provides a standard interface for
operator interfaces (LIGO uses the ``MEDM'' GUI tool) and supervisory
control, such as \guardian, the Advanced LIGO automation
platform~\cite{guardian}.

\subsection{Software Architecture}

The \advligorts\ software consists of three primary components: a
patched version of the standard Linux kernel, a \textit{real-time code
  generator} (RCG), and a suite of data acquisition daemons (DAQD).

The Linux kernel patch, which is simple and small, allows loadable
kernel modules to request that the kernel remove a specific CPU from
the normal linux process scheduler and hand it over for exclusive use
by the module code.  The kernel module can then have uninterruptible
use of the CPU at full rate.

Typically \advligorts\ programs are drawn as signal flow diagrams in
\MATLAB\ Simulink.  The RCG takes a Simulink model as input, parses it
to extract the signal flow graph, and outputs C code that can execute
the same signal flow logic.  The RCG adds wrapper code for process
synchronization and inter-process communication, then compiles the
resulting code into an \textit{RTS} linux kernel module that can be
immediately inserted into the running kernel.  The RTS module is
completely self sufficient at runtime, requiring no services from the
kernel or the rest of the operating system.  A special RTS module
called an \textit{input-output processor} (IOP) handles all I/O
processing, reading data from the ADCs and passing it to other RTS
modules via a shared memory interface, and writing output data passed
from RTS modules to the DACs.  The IOP processes runs at 65536~Hz, and
normal RTS modules can run at any power of two less than that.

Each RTS module employs various user space processes to handle
user-facing IO tasks.  An \EPICS\ \textit{input/output controller}
(IOC) process exposes writable parameters of the controller logic to
an \EPICS\ message passing system~\cite{epics}.  The \texttt{awgtpman}
process handles the activation of temporary test point outputs and the
processing and insertion of signal injections sent over the network
(an interface protocol referred to as ``AWG'').  Each of these user
space processes uses shared memory to exchange data with their
respective RTS kernel space processes.  Interconnection between RTS
processes is achieved either via shared memory for RTS modules in the
same kernel, or via a Dolphin PCIe shared memory
network~\cite{dolphin} for RTS modules on different hosts.

Each front end host computer also runs two processes that collect data
from the RTS processes via shared memory (\texttt{local\_dc}) and send
the data over the network in 1/16th-second chunks
(\texttt{data\_xmit}).  The data acquisition chain receives data from
all front ends and collates them into 1/16th-second blocks that are
consumed by the DAQD processes which write the data to disk, serve it
over the network, or forward it to other streaming processes.  In the
production environment, the DAQD processes are spread across multiple
machines to reduce resource competition.

\subsection{Software Functionality}

\advligorts\ leverages \MATLAB\ Simulink to provide a powerful
graphical user interface for users to draw signal flow graphs that can
be turned into real-time code (see Section~\ref{sec:examples}).  Users
can drag and drop parts from an extensive parts library and easily
connect them together.  The parts library includes parts for all
supported I/O interfaces, most math operations, matrices, logic gates,
switches, oscillators, demodulation, etc., allowing for essentially
arbitrary signal processing.

Most of the parameters of all parts are exposed as process variables
through the \EPICS\ interface.  Process variables can take the form of
floats, integers, or strings, and are used to control switches, set
signal gains and matrix coefficients, etc.  The \EPICS\ interface also
provides monitor test points for the signals passing through the
system.  Many tools exist to interface with \EPICS, including python
libraries, command line tools, and various graphical programs for
creating operator interface screens.

One of the most important parts is the filter module.  Filter modules
hold a bank of ten addressable digital filters, each with up to 20
poles and zeros, implemented as cascaded second-order-sections.  A
companion filter design tool called \texttt{foton}\ allows users to
design filters from scratch, or draw from a library of common filter
functions.  Once loaded, individual filters can be engaged or
disengaged in situ via a single command from the \EPICS\ interface.
The filter modules also include excitation inputs, test points,
switches for input and output engagement, an additive offset, and a
multiplicative gain.

The RCG also allows users to define their own \texttt{C} code
functions that can be dropped anywhere in the signal flow graph to
perform arbitrary signal manipulations.

The data acquisition daemon provides a Network Data Service (``NDS'')
which can be used to access all data acquired by the system in
real-time, or to access archival data that has been stored to disk.
This service also allows users to request real-time data from virtual
test points in the system.  The NDS test point interface and the AWG
signal injection interface together provide a critical interface for
real-time characterization of the system not readily available on
other digital signal processing systems.  Various supporting
applications give scientists the ability to plot the data in both the
time and frequency domains, and to make various excition-based
measurements of the system.

\section{Illustrative Examples}
\label{sec:examples}

Each Advanced LIGO interferometer employs over 120 RTS models, spread
across nearly three dozen front end computers.  The data acquisition
systems process roughly 10k ``fast'' channels (with sample rates from
512 - 16k~Hz) and nearly 300k ``slow'' channels (16~Hz, EPICS process
variables).

\begin{figure}
  \centering
  \begin{subfigure}[b]{0.9\textwidth}
    \fbox{\includegraphics[width=\textwidth]{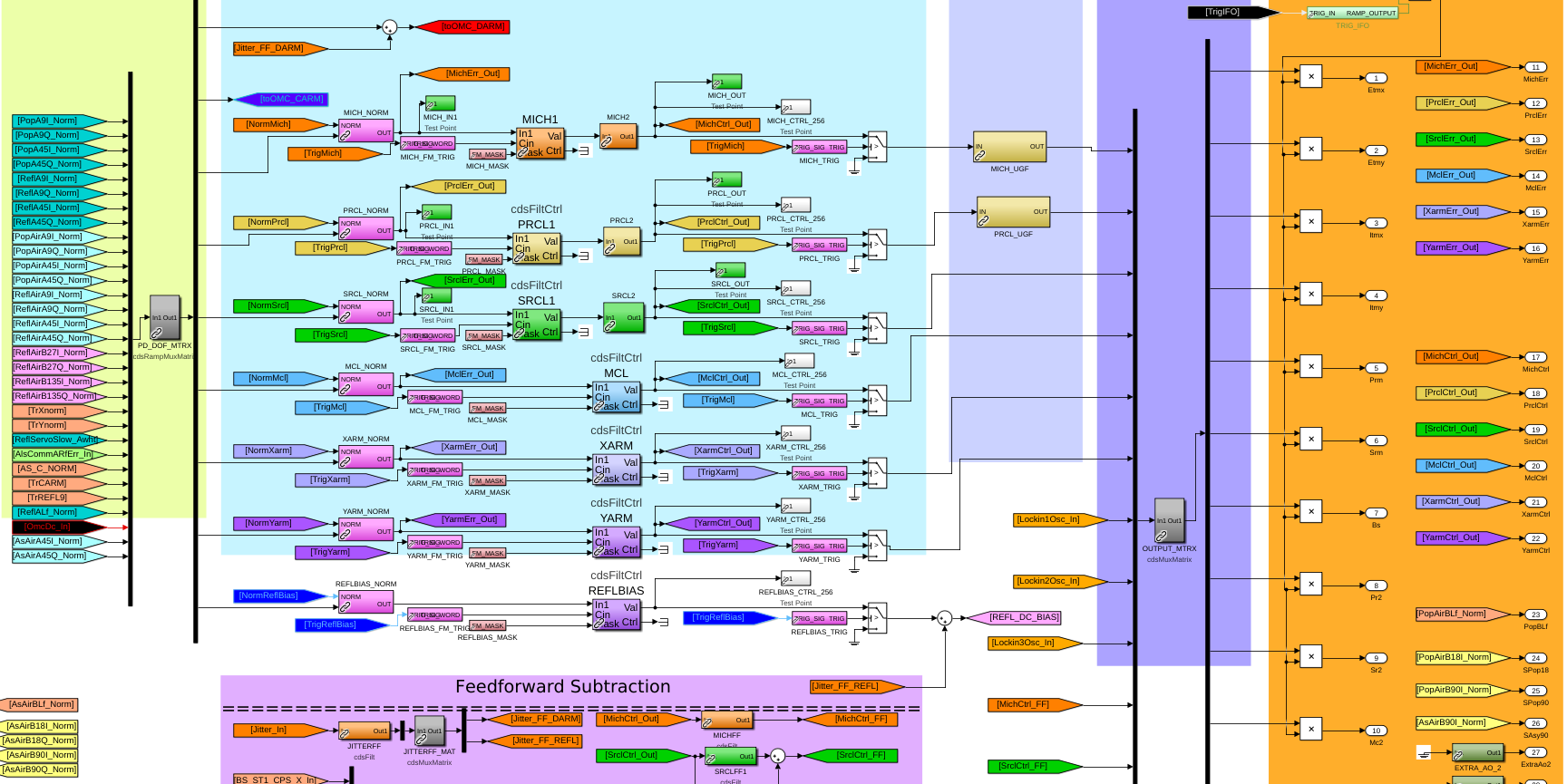}}
    \caption{Excerpt of Simulink code for the LSC real-time
      controller}
    \label{fig:lsc:simulink}
  \end{subfigure}

  \begin{subfigure}[b]{0.9\textwidth}
    \fbox{\includegraphics[width=\textwidth]{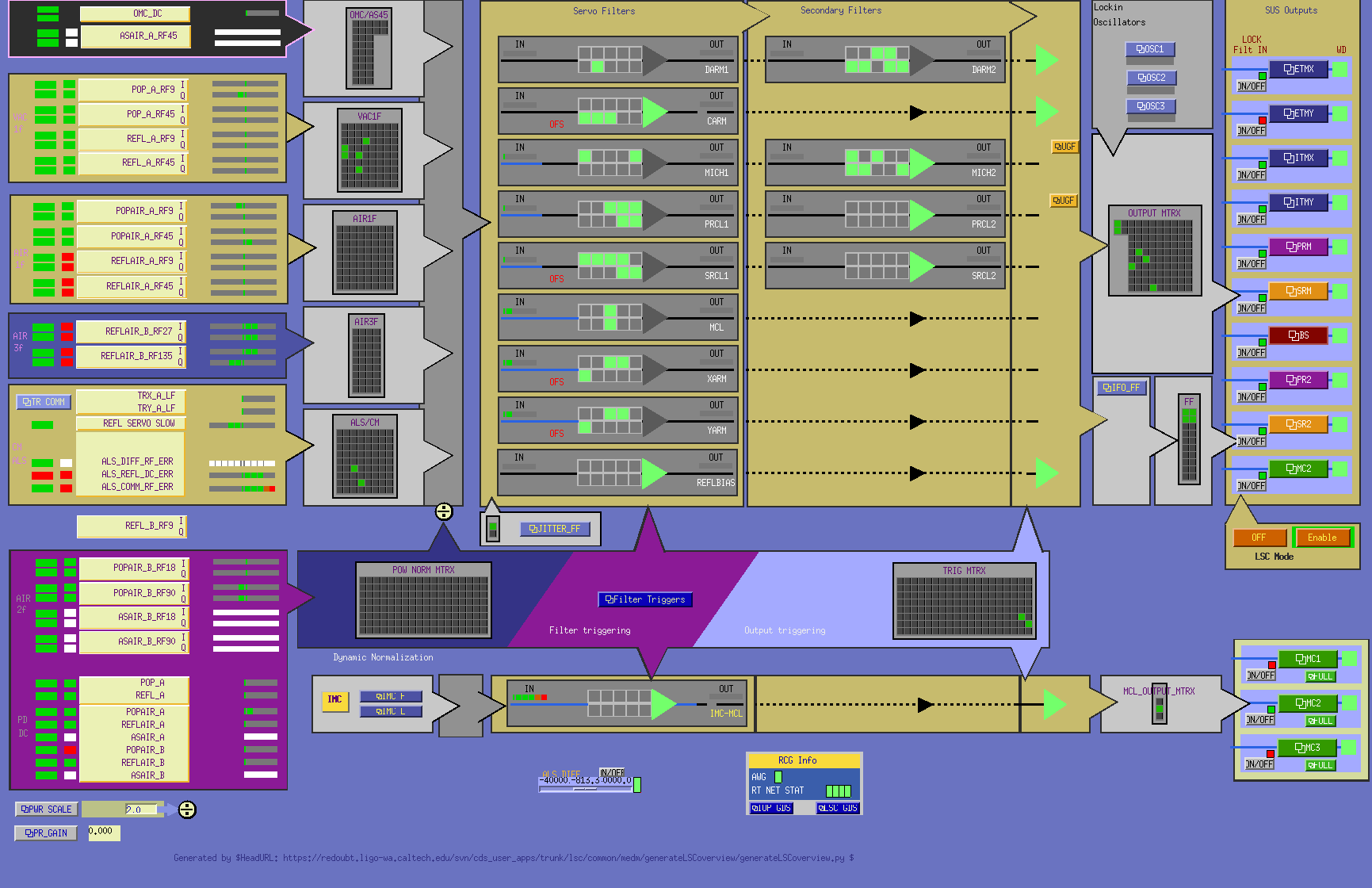}}
    \caption{Screenshot of the operator interface for LSC real-time
      controller}
    \label{fig:lsc:medm}
  \end{subfigure}
\caption{A small excerpt of Simulink code for the length sensing and
    control (LSC) subsystem, and a screen shot of the corresponding
    operator interface.  See text for description of the components
    and structure of the code.}
  \label{fig:lsc}
\end{figure}

Figure~\ref{fig:lsc:simulink} shows an excerpt of the Simulink code
used to program the real-time controller for the LIGO length sensing
and control subsystem which controls the lengths of all the various
optical cavities in the Advanced LIGO interferometers.  This model
runs at 16k~Hz, taking as input the RF-demodulated photodetector error
signals.  It's outputs are fed over the Dolphin network to separate
RTS models that control the suspended optics.

Signal flow is drawn from left to right, with outputs for this block
at the far right.  The diagram is less busy that it might otherwise be
because of the liberal use of GOTO/FROM connections that eliminate the
need for drawn lines connecting the output of one block to the input
of another.  Filter modules are labeled ``\texttt{cdsFiltCtrl}''.  At
the left of the diagram, a gray box represents a matrix (labeled
``PD\_DOF\_MTRX'') that handles ``rotation'' of sensor input signals
into the canonical longitudinal degree of freedom basis for the
interferometers.  Near the right another matrix (``OUTPUT\_MTRX'')
handles rotation of signals in the degree of freedom basis into the
basis of output suspension drive actuators.

\section{Impact}

Construction of the Advanced LIGO detectors was completed in 2013.
The first gravitational wave was detected in September
2015~\cite{doi:10.1103/physrevlett.116.061102}.  The flexibility,
modularity, and ease of programming allowed by the \advligorts\ system
was critical for the rapid commissioning of the detectors down to
unprecedented sensitivities.  The ongoing success of LIGO (there have
been dozens more confirmed detections since
GW150914~\cite{doi:10.1103/physrevx.9.031040}) is a testament to the
robustness of this system.

\begin{figure}
  \includegraphics[width=\columnwidth]{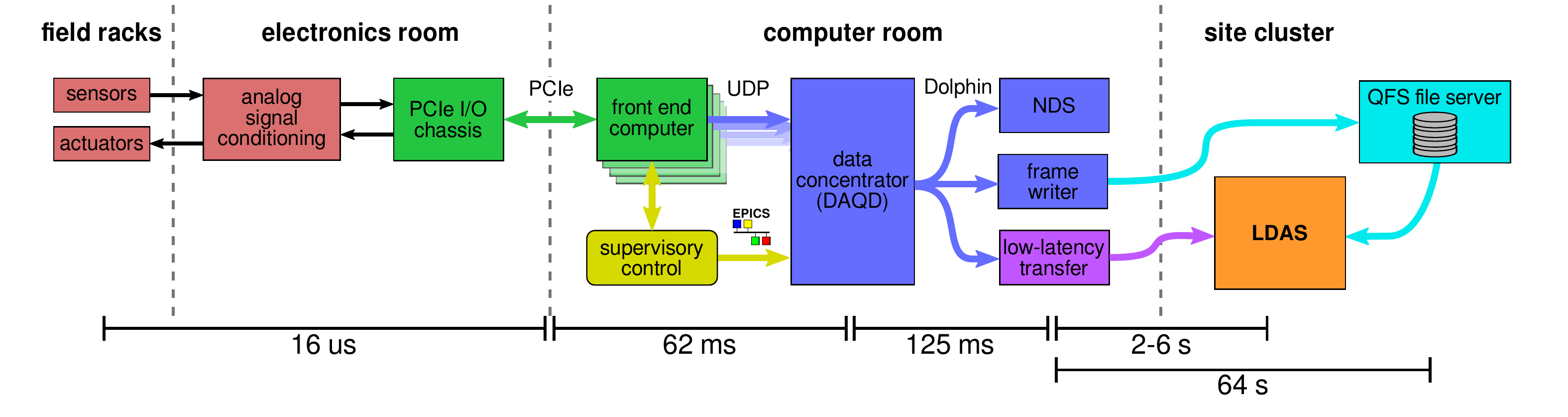}
  \caption{Advanced LIGO's data processing pipeline.  Times shown at
    the bottom are sample latencies.  PCIe, UDP, and Dolphin refer to
    the transport fabrics used.  The gravitational wave searches are
    conducted on the LIGO ``LDAS'' computer clusters, which receive
    data from the observatories within seconds.}
  \label{fig:daq}
\end{figure}

In August 2017, LIGO made the first ever detection of gravitational
waves from a binary neutron star
merger~\cite{doi:10.1103/physrevlett.119.161101}.  With help from the
Virgo detector~\cite{doi:10.1088/0264-9381/32/2/024001}, the merger
was localized to an area in the sky of roughly
30~deg$^2$~\cite{doi:10.3847/2041-8213/aa91c9}.  Within hours of
detection and localization, alerts had been sent to the astronomical
community and dozens of electromagnetic observatories across the
world, who then made the first observation of a gamma-ray burst
emitting kilonova~\cite{doi:10.3847/2041-8213/aa91c9}.  These
observations were made possible by a data acquisition system that was
able to reliably deliver strain data from the detectors to the search
pipelines within a matter of seconds (see Figure~\ref{fig:daq}).  The
\advligorts\ data acquisition system is the source of this entire
pipeline.

Advanced LIGO has already undergone one significant upgrade,
installing a squeezed light source which reduces quantum noise at high
frequencies~\cite{doi:10.1103/physrevlett.123.231107}.  And the more
significant ``A+'' upgrade will commence at the end of the current
observing run, adding another new subsystem to shape the quantum noise
suppression from the squeezed light source, and further increasing the
sensitivity of the current detectors to less than
$10^{-22}\ \mathrm{m}/\sqrt{\mathrm{Hz}}$.  These upgrades are enabled
by the flexibility and extensibility of \advligorts\ framework.

With the success of LIGO, use of the \advligorts\ system is spreading
throughout the gravitational wave community.  The Japanese project
KAGRA~\cite{doi:10.1103/physrevd.88.043007} has adopted the full
Advanced LIGO digital control and data acquisition system for control
of their underground gravitational wave detector.  \advligorts\ is
also used in the GEO 600 project, the Caltech 40m prototype, the MIT
LASTI prototype, the AEI 10m prototype, as well as in dozens of
smaller laboratories around the world to control a variety of
table-top opto-mechanical experiments.

\section{Conclusions and future development}

LIGO is expected to operate gravitational wave observatories for at
least the next two decades.  In that time, a third identical detector
is expected to be completed in
India~\cite{doi:10.1007/s41114-018-0012-9}.  Beyond this, a new
generation of ground-based detectors is being planned to extend our
reach out to the edge of the observable
universe~\cite{arxiv:2001.11173, doi:10.1088/1361-6382/aa51f4,
  doi:10.1088/0264-9381/27/19/194002}.  These new detectors will have
even more complex controls challenges, and the \advligorts\ system
will need to continue to evolved and adapt to meet their needs.

Development is ongoing to improve the usability and extend the
functionality of \advligorts.  A key development track aims to see if
the need for a specially-patched linux kernel can be eliminated, by
leveraging new features in the standard linux kernel that allow for
CPU-isolation for privileged processes.  Developers have also been
experimenting with running RTS processes in user space.  Unprivileged
user space execution would not be for real-time operation, but could
allow for running code faster than real-time for simulation or testing
purposes.

A project is also underway to explore the possibility of executing
neural networks within RTS processes.  Neural networks trained on
simulations and archived data could be ported into RTS processes to
enable machine learning experimental control.  There is potential for
training and updating these networks in real-time, for more
sophisticated reinforcement learning applications.

The data acquisition pipeline is being improved to increase the
throughput and accessibility of larger amounts of data.  Currently,
only a small subset of the full channel data is available in low
latency, limiting the amount of data characterization that can be done
for the low-latency search pipelines.  It should be possible to dump
the full data pipe into the local LDAS clusters at a fraction of the
current time, decreasing latencies for multi-messenger searches and
potentially leading to more ground-breaking discoveries.

\section{Conflict of Interest}

We wish to confirm that there are no known conflicts of interest
associated with this publication and there has been no significant
financial support for this work that could have influenced its
outcome.

\section*{Acknowledgments}

LIGO was constructed by the California Institute of Technology and
Massachusetts Institute of Technology with funding from the National
Science Foundation and operates under Grant No. PHY-0757058. Advanced
LIGO was built under award PHY-0823459.

\bibliography{bib,extra}
\bibliographystyle{habbrv}

\end{document}